\documentclass[aps,twocolumn,prd,preprintnumbers, 10pt]{revtex4-2}
\usepackage{amsmath}
\usepackage{graphicx}
\usepackage{bbm}
\usepackage{amssymb}
\usepackage{mathrsfs}
\usepackage{slashed}
\usepackage{cancel}
\usepackage[linktocpage=true,
colorlinks=true,
urlcolor=blue,
anchorcolor=blue,
citecolor=blue,
filecolor=blue,
linkcolor=blue,
menucolor=blue
]{hyperref}
\usepackage[normalem]{ulem}
\usepackage{soul,xcolor}

\newcommand{\ii}{\mathrm{i}}
\newcommand{\R}{\mathbb{R}}
\newcommand{\ddiff}{\mathrm{d}}
\newcommand{\vol}{\mathrm{vol}}

\newcommand{\hook}{\mathbin{\rule[.2ex]{.4em}{.03em}\rule[.2ex]{.03em}{.9ex}}}

\newcommand{\Fgrav}{F_{\mathrm{grav}}}

\newcommand\cF{\mathcal{F}}

\def\nn{\nonumber}

\newcommand{\pmu}{{\kappa}}

\newcommand{\xinew}{\zeta}
\newcommand{\dd}{\mathrm{d}}

\usepackage{color}

\def\nn{\nonumber}

\begin{document}

\setstcolor{red}

\title{Localization in supergravity}

\author{James Sparks}

\affiliation{Mathematical Institute, University of Oxford, Woodstock Road, Oxford, OX2 6GG, U.K.}

\begin{abstract}
\noindent  We give an introduction to equivariant localization in supergravity, focusing 
on the application to four-dimensional theories and supersymmetric black holes.

\vskip 0.3cm

\noindent \emph{Invited contribution to the proceedings book ``Modern topics in mathematical physics''}

\end{abstract}

\maketitle

\section{Introduction}\label{sec:intro}

Supersymmetric solutions of supergravity theories have played a central role 
in string theory and quantum gravity. Supersymmetry provides additional 
structure, with the analytic tools this brings allowing for precise calculations.
Notable examples include the microscopic interpretation of entropy for supersymmetric black holes, and the AdS/CFT correspondence which relates
quantum gravity to a dual conformal field theory on the boundary of spacetime.
The last decade has seen these two subjects 
brought together for asymptotically anti-de Sitter (AdS) black holes, starting with the work of \cite{Benini:2015eyy, Benini:2016rke}.
In particular, new approaches have been developed for computing various protected physical observables exactly in supersymmetric field theories \cite{Pestun:2016zxk}. 
Many of these results have been matched to a semi-classical analysis 
in gravity, confirming and sharpening our understanding of holography, but much remains to be understood. 

The simplest observable one can consider in a theory is the partition function $Z$, 
or equivalently the free energy $F=-\log Z$. If quantum gravity 
is described by some version of a path integral, then in the leading 
semi-classical limit $F=I_{\text{on-shell}}$ is identified with the on-shell action of 
Euclidean solutions to the classical gravitational equations of motion~\cite{Gibbons:1976ue}. 
Understanding which saddle points contribute, and evaluating those contributions, is a key problem. 
The Einstein equations are coupled non-linear
partial differential equations, and finding solutions in closed form usually relies on a high degree of symmetry. However, we are not interested in 
finding the solutions {\it per se}, but rather in evaluating their contribution 
to the partition function.

Reference \cite{BenettiGenolini:2023kxp} introduced a new approach: \emph{equivariant localization in supergravity}. This is a mathematical structure
 that allows one to compute many observables for supersymmetric solutions,
without solving the supergravity equations.
Supergravity localization has been applied to compute on-shell actions for theories 
in four \cite{BenettiGenolini:2023kxp, BenettiGenolini:2024xeo, BenettiGenolini:2024hyd, BenettiGenolini:2024lbj, Hristov:2024cgj}, five \cite{Cassani:2024kjn, Colombo:2025ihp, BenettiGenolini:2025icr, Colombo:2025yqy, Park:2025fon} (with variants of the supergravity localization approach) and 
six dimensions \cite{Couzens:2025ghx}. The 
structure is very general, allowing one to also compute  black hole
entropies, central charges, scaling dimensions of operators, and other quantities 
\cite{BenettiGenolini:2024kyy, BenettiGenolini:2023ndb, BenettiGenolini:2023yfe, Suh:2024asy, Couzens:2024vbn, Couzens:2025nxw, Couzens:2026qne, Martelli:2023oqk, Colombo:2023fhu, Cassia:2025aus, Cassia:2025jkr}. 
The main observation of \cite{BenettiGenolini:2023kxp}
is that supersymmetric supergravity solutions are equipped with a 
canonical set of polyforms $\Phi$, {\it i.e.} formal sums of differential forms of 
different degrees. These forms are closed $\ddiff_\xi \Phi=0$ under the operator 
$\ddiff_\xi \equiv \ddiff - \xi \hook\mskip2mu$, where $\ddiff$ is the usual exterior 
derivative and $\xi\hook$ denotes contraction with 
the vector field $\xi$. Here $\xi$ is a Killing vector field that is also defined 
 canonically for supersymmetric solutions, that we call the \emph{supersymmetric Killing vector}. This generates a symmetry of the solutions, and 
the set of fixed points where $\xi=0$ will play an important role. The cohomology 
of the operator  $\ddiff_\xi$ is called \emph{equivariant cohomology}, and 
the observables 
we compute represent equivariant cohomology classes.

The following general structure has emerged by studying supergravity theories in diverse dimensions, coupled to 
different forms of matter, and for various compactifications of string theory and M-theory:
\begin{enumerate}
\item There typically exist many equivariantly closed 
polyforms $\Phi$ in a given supergravity theory, {\it i.e.} satisfying $\ddiff_\xi\Phi=0$.
\item Interesting physical quantities arise as integrals of 
these forms, which may be evaluated using the Berline--Vergne--Atiyah--Bott (BVAB) fixed point formula \cite{BV:1982, Atiyah:1984px}. 
\item The quantities that appear in these localized expressions 
may be evaluated without knowing the explicit form of solutions.
\end{enumerate}
This last point is key: the expressions we find are always 
``equivariant topological invariants,'' depending only on weights of the vector field 
$\xi$ at fixed points, and certain global data/topological invariants. 
This points to the existence of a protected sector in supergravity theories, 
with the observables  being captured by equivariant cohomology. 

In this article we give an introduction to supergravity localization in the simplest setting, focusing on $D=4$, $\mathcal{N}=2$ minimal gauged 
supergravity and its coupling to vector multiplet matter. This allows 
us to develop the general theory and explain key concepts and results, 
 illustrating with simple examples that include classes of black hole 
solutions. 

\section{$D=4$ minimal gauged supergravity}\label{sec:minimal}

\subsection*{The theory}

The bosonic field content of 
$D=4$, $\mathcal{N}=2$ minimal gauged supergravity \cite{Freedman:1976aw} 
 is simply Einstein--Maxwell theory with a negative cosmological constant. The bulk action in Euclidean signature is 
\begin{align}\label{bulkI}
I_{\text{bulk}} =  - \frac{1}{16\pi G_4} \int_{M_4} (R+6-F^2)\, \vol_g\, . 
\end{align}
Here $G_4$ is the Newton constant, $R$ is the Ricci scalar, 
and $F=\ddiff A$ is the Maxwell field strength with $F^2\equiv F_{\mu\nu}F^{\mu\nu}$, $\mu,\nu=1,\ldots,4$. We have normalized the cosmological constant to $\Lambda=-3$, and denote
Riemannian volume forms as $\vol_g=\sqrt{\det g}\, \ddiff^4 x$. 
The action \eqref{bulkI} leads to the  Einstein--Maxwell equations
\begin{align}
\label{EOM}
 R_{\mu\nu} + 3g_{\mu\nu} -2\left(F_{\mu\rho}{F_{\nu}}^{\rho} - \tfrac{1}{4}F^2g_{\mu\nu}\right)  &= 0 \, ,\nonumber\\
\ddiff * F  & = 0\, .
\end{align}
Here we take the (not entirely standard) convention that a $p$-form $\alpha$ has Hodge dual the
$(4-p)$-form
$(*\mskip2mu\alpha)_{\mu_1\cdots\mu_{4-p}}\equiv\frac{1}{p!}\varepsilon_{\mu_1\cdots\mu_{4-p}}^{\ \ \ \ \ \ \ \ \ \ \nu_1\cdots\nu_p}\alpha_{\nu_1\cdots \nu_p}$. 

A solution to the equations of motion \eqref{EOM} is  by definition supersymmetric if the supersymmetry variations of all fields is zero. For a bosonic solution with all fermions set to zero, the only 
non-trivial supersymmetry variation is that for the  gravitino field~$\psi_\mu$:
\begin{align}
\label{SUSY}
\delta_\epsilon \psi_\mu = \Big(\nabla_{\mu} - \ii A_{\mu} + \frac{1}{2}\gamma_{\mu} + \frac{\ii}{4}F_{\nu\rho}\gamma^{\nu\rho}\gamma_{\mu}\Big)\epsilon = 0 \, .
\end{align}
Here $\epsilon$ is a Dirac spinor, $\nabla_\mu=\partial_\mu + \tfrac{1}{4}\omega_\mu^{\ ab}\gamma_{ab}$ is the spin connection with respect to an orthonormal frame $e^a_\mu$, $a=1,\ldots,4$, and  the Hermitian gamma matrices $\gamma_\mu$ generate the Clifford algebra Cliff$(4,0)$, so $\{\gamma_\mu,\gamma_\nu\} = 2g_{\mu\nu}\, \mathbbm{1}$. 
Notice from the second term in \eqref{SUSY} that the spinor is charged 
under the gauge field $A$ -- this gauging of the R-symmetry that rotates the spinor 
$\epsilon$ by a phase is what makes it a ``gauged supergravity.''  
 Equation \eqref{SUSY} is a form of generalized Killing spinor equation. 
If a  (not identically zero) solution $\epsilon$ exists, 
it automatically follows that the metric and gauge field satisfy 
the equations of motion \eqref{EOM} \cite{Genolini:2021urf}.

For simplicity we take the metric and gauge field to be real in what follows. More generally 
in Euclidean signature one should allow for complex fields, in the spirit of 
finding saddle points to the gravitational path integral. This doubling of the number of real degrees of freedom implies that, with supersymmetry, also the spinor $\epsilon$ 
and its complex conjugate are independent. See \cite{Freedman:2013oja,  BenettiGenolini:2024lbj} for further discussion. 

One reason to be interested in 
this theory is that it is a ``consistent truncation'' of 
eleven-dimensional supergravity on a general class of internal spaces 
called Sasaki--Einstein manifolds
\cite{Gauntlett:2007ma}. This means that any supersymmetric solution 
in $D=4$ uplifts to a supersymmetric solution in $D=11$, where 
we have a more precise (microscopic) understanding of  holography.  
For example, taking the internal space to be 
 the seven-sphere $S^7$ means that the 
 four-dimensional gravitational theory is holographically dual to the 
three-dimensional Aharony--Bergman--Jafferis--Maldacena (ABJM) theory 
\cite{Aharony:2008ug}. This is a Chern--Simons theory coupled to matter, 
with gauge group $U(N)\times U(N)$. The gravitational dual 
is described by classical supergravity in the
 large $N$ limit: at leading order, the 
Newton constant in \eqref{bulkI} is related to $N$ by 
$\frac{1}{16\pi G_4} = \frac{\sqrt{2}}{24\pi}N^{3/2}$.  
We will focus on localization in gravity in this article, rather than 
trying to review holography, but this is an important motivation 
for what follows.

\subsection*{Bilinear forms}

A standard approach to studying supersymmetric solutions is to make an ansatz 
for the metric, gauge field and spinor $\epsilon$, typically imposing symmetries in order to 
reduce the partial differential equations (PDEs) to ordinary differential equations (ODEs). 
While many 
 interesting solutions have been constructed in this way, we take a more general and indirect approach. 

Following \cite{Caldarelli:2003pb, BenettiGenolini:2019jdz}, from the Killing spinor $\epsilon$ one may
construct the real bilinears
\begin{align}\label{4dbilinears}
 S & \equiv  \bar{\epsilon}\epsilon\, , \quad   P   \equiv \bar{\epsilon}\gamma_5 \epsilon\, , \quad  \xi^\flat_\mu  \equiv -\ii \bar{\epsilon}\gamma_\mu\gamma_5\epsilon\, , \quad  \nonumber\\
K_\mu &  \equiv \bar{\epsilon}\gamma_\mu\epsilon\, , \quad  U_{\mu\nu}    \equiv \ii \bar{\epsilon}\gamma_{\mu\nu}\epsilon\, .
\end{align}
Here $\gamma_5\equiv \gamma_{1234}$ satisfies $\gamma_5^2=1$, and $\bar\epsilon\equiv \epsilon^\dagger$ 
denotes the Hermitian conjugate of $\epsilon$. $S$ and $P$ are scalar functions, 
$\xi^\flat_\mu$ and $K_\mu$ are one-forms, and $U_{\mu\nu}$ is a two-form. 
We denote by $\xi^\mu=g^{\mu\nu}\xi^\flat_\nu$ the vector field dual to the one-form $\xi^\flat$.

Using the Killing spinor equation one can take covariant 
derivatives $\nabla_\rho$ of the objects in \eqref{4dbilinears}. 
Here one uses \eqref{SUSY} and its Hermitian conjugate to substitute for 
$\nabla_\rho\epsilon$, $\nabla_\rho \bar\epsilon$ in terms of  the other terms in 
\eqref{SUSY}. 
In particular, in this way one can
check that $\xi^\mu$ is a Killing vector, $\nabla_{(\mu}\xi^\flat_{\nu)}=0$, while
skew-symmetrizing over indices leads to the one-form, two-form and three-form differential equations:
\begin{align}\label{dforms}
\ddiff S & = - K + \xi\hook * \mskip3mu F\, , \quad  \ddiff P = - \xi\hook F\, ,\nonumber\\
\ddiff \xi^\flat & = -2 (*\mskip3mu U + S*F+PF)\, ,\nonumber\\
\ddiff U & = -3 *\xi^\flat +F\wedge K\, .
\end{align}
Here  $(\xi\hook F)_\mu \equiv \xi^{\nu}F_{\nu\mu}$. 
Notice that using the Cartan formula we have $\mathcal{L}_\xi F = 
\xi \hook (\ddiff F) + \ddiff(\xi \hook F) = 0$, where we have used the Bianchi identity $\ddiff F=0$ and 
the second equation in \eqref{dforms}. Thus $\mathcal{L}_\xi g_{\mu\nu}=0=\mathcal{L}_\xi F_{\mu\nu}$ and $\xi$ generates a symmetry of the full solution. 
We call $\xi$ the \emph{supersymmetric Killing vector}.

One can similarly construct further bilinear forms, and compute their derivatives using 
\eqref{SUSY}, but these will not be needed in what follows. 

\subsection*{Equivariantly closed forms}

The previous subsection follows 
 the ``$G$-structure'' approach 
to analysing supersymmetric supergravity solutions: one recasts the first order Killing spinor equation~\eqref{SUSY} in terms of
first order differential equations for bilinear forms,  of which \eqref{dforms} are a subset. For a nice review, see \cite{Gauntlett:2005bn}. However, 
 the next step was introduced much more recently in \cite{BenettiGenolini:2023kxp}.

Given the existence of a canonical vector $\xi$ generating a symmetry, and differential forms, it is natural to introduce 
the equivariant exterior derivative \cite{BV:1982, Witten:1982im}
\begin{align}
\ddiff_\xi \equiv \ddiff - \xi \hook\, .
\end{align}
This acts on formal sums of differential forms of mixed degree, {\it i.e.} polyforms. 
The exterior derivative $\ddiff$ raises the degree of a form by one, while contraction 
with the vector $\xi$ lowers the degree by one. The Cartan formula 
gives $\ddiff_\xi^2 = - \mathcal{L}_\xi$. This means that acting on polyforms 
that are invariant under the Lie derivative $\mathcal{L}_\xi$, the operator 
$\ddiff_\xi$ is nilpotent. The corresponding cohomology, generalizing the de Rham cohomology when $\xi=0$, is called equivariant cohomology  \cite{BV:1982}. 

Given the equations \eqref{dforms}, a little inspiration leads one to define
 the polyforms
\begin{align}\label{Phis}
\Phi^F & = \Phi^F_2 + \Phi^F_0  \equiv F - P\, , \nonumber\\
\Phi  & =\Phi_4+\Phi_2+\Phi_0   \\ & \equiv (3\mskip2mu \vol_g + F\wedge *\mskip2mu F)+(U+SF-P*F)-(SP)\, .\nonumber
\end{align}
Here the subscripts denote degrees of the forms, so {\it e.g.} 
$\Phi^F_2=F$ is a two-form, $\Phi^F_0=-P$ is a function. Using~\eqref{dforms} one checks that 
\begin{align}
\ddiff_\xi \Phi^F = 0 = \ddiff_\xi \Phi\, ,
\end{align}
are both equivariantly closed. For example, we compute
\begin{align}
\ddiff_\xi \Phi^F = \ddiff F -\xi \hook F - \ddiff P = 0\, ,
\end{align}
using the Bianchi identity and the second equation in~\eqref{dforms}. 
Checking $\ddiff_\xi \Phi=0$ is only slightly more work, and requires using the algebraic (Fierz) one-form identity 
$\xi\hook \mskip2mu U = PK$. 

The top form of $\Phi^F$ is the gauge field curvature $F$, and integrating it through a two-dimensional submanifold $\Sigma\subset M_4$ by definition gives the quantized magnetic flux $\frac{1}{4\pi}\int_\Sigma F$  through $\Sigma$.
On the other hand, taking  the trace of the Einstein equation \eqref{EOM} gives $R=-12$. 
Substituting this into the bulk action \eqref{bulkI}, and noting 
$F^2\mskip2mu \vol_g = 2F\wedge *\mskip2mu F$, we see that the 
bulk on-shell 
action~\eqref{bulkI} is given by
\begin{align}\label{bulkIonshell}
I_{\text{bulk, on-shell}} = \left[\frac{1}{(2\pi)^2}\int_{M_4} \Phi_4\right]\frac{\pi}{2G_4} \, .
\end{align}
Here $\Phi_4$ is precisely the top form of the equivariantly closed 
polyform $\Phi$ in \eqref{Phis}. 

\subsection*{Fixed point formula}

Having found some equivariantly closed forms, the Berline--Vergne--Atiyah--Bott (BVAB) formula \cite{BV:1982, Atiyah:1984px} allows one to express integrals of these forms 
in terms of quantities evaluated at the fixed point set $\mathcal{F}\equiv\{\xi=0\}\subset M$. Here recall that a vector field $\xi$ generates a flow on a manifold $M$, where one moves points of $M$ along the integral curves of $\xi$. The fixed points of this flow are precisely the points where $\xi=0$.

Here we give an account in dimensions $D=2, 4$, before applying to 
supergravity in the next subsection.

\vskip 0.25cm

\noindent {$\mathbf{D=2}$}: As a simple example, consider the two-sphere $M_2=S^2$ equipped with the vector field $\xi = b\mskip2mu \partial_{\varphi}$. Here $b\neq 0$ is a 
constant parameter and $(\theta,\varphi)$ are spherical polar coordinates, so that 
$\xi$ rotates $S^2$ about its axis.
The fixed point set  $\mathcal{F}$ consists of two points: 
the north pole ($\theta=0$) and south pole ($\theta=\pi$).  
 
If $\Phi=\Phi_2+\Phi_0$ is equivariantly closed, $\ddiff_\xi\Phi=0$, the BVAB formula states
\begin{align}\label{2dBVAB}
\int_{S^2}\Phi_2 & = \sum_{\text{fixed points}\, p}\frac{2\pi}{b_p} \Phi_0|_p \nonumber \\
& =  \frac{2\pi}{b}\Phi_0|_{\theta=0} - \frac{2\pi}{b} \Phi_0|_{\theta=\pi}\, .
\end{align}
Here $\Phi_0|_p$ denotes the function $\Phi_0$ evaluated at the point $p$, 
while $b_p$ denotes the \emph{weight} of $\xi$ at $p$. 
A perennial subtlety 
when using the BVAB formula is determining signs of weights consistently: at the north pole 
$b_p=b$, where $\partial_\varphi$ rotates the tangent space with weight $+1$ (anti-clockwise), 
while at the south pole $b_p=-b$, where $\partial_\varphi$ rotates the tangent space with weight $-1$ (clockwise).  The BVAB formula \eqref{2dBVAB} 
relates the integral of the top form $\Phi_2$ to the lower degree form $\Phi_0$ 
and weights at the fixed points.

For example, we can take $\Phi=\sin\theta\mskip2mu \ddiff\theta\wedge \ddiff\varphi + 
b \cos\theta+c$. Note the arbitrary additive constant $c$. Then from \eqref{2dBVAB} the volume of the round metric on $S^2$ is
\begin{align}
\int_{S^2}\Phi_2 = \frac{2\pi}{b}\, (b+c) - \frac{2\pi}{b}\, (-b+c) = 4\pi\, ,
\end{align}
where $c$ has necessarily dropped out. 

\vskip 0.25cm

\noindent {$\mathbf{D=4}$}: Now let $M_4$ be a compact four-manifold without boundary, 
and $\Phi=\Phi_4+\Phi_2+\Phi_0$ be an equivariantly closed polyform, $\ddiff_\xi \Phi=0$. The BVAB formula in this case gives
\begin{align}\label{4dBVAB}
\int_{M_4}\Phi_4 & = \sum_{\text{nuts}\, p}\frac{(2\pi)^2}{b_{1,p} b_{2,p}}\Phi_0|_p \nonumber\\ 
& + 
\sum_{\text{bolts}\, \Sigma}  \frac{2\pi}{b_\Sigma} \int_\Sigma \Big[\Phi_2 - \frac{2\pi}{b_\Sigma}\Phi_0\, c_1(N)\Big]\, .
\end{align}
On a four-manifold the fixed point set $\mathcal{F}$ consists of isolated fixed points $p$
(called \emph{nuts}) and fixed two-dimensional surfaces $\Sigma$ (called \emph{bolts}) \cite{Gibbons:1979xm}:
\begin{itemize}
\item 
At an isolated nut $p$, the linear action of $\xi$ on the tangent space splits $TM_4 |_p=\R^4=\R^2\oplus\R^2$, with weights $b_i$ on each factor $\R^2_i$, $i=1,2$. One can introduce plane polar coordinates for each $\R^2_i$ and write 
$\xi = b_1\partial_{\varphi_1}+b_2\partial_{\varphi_2}$ near the fixed point. 
Equivalently, in an orthonormal frame at the nut we have
\begin{align}\label{nutweight}
\frac{1}{2}\ddiff\xi^\flat |_p = \begin{pmatrix} 0 & b_1 & 0 & 0 \\ -b_1 & 0 & 0 & 0\\
0 & 0 & 0& b_2\\ 0 & 0 & -b_2& 0\end{pmatrix}\, ,
\end{align}
so that  $\ii b_1$, $\ii b_2$ are the skew eigenvalues of this matrix. 
Note that which weight is which is a convention, as is the orientation of $\R^2_i$, provided the orientation of $\R^4$ agrees with the orientation one uses to integrate. 
In practice this means formulae should be invariant under $(b_1,b_2)\leftrightarrow (b_2,b_1)$, $(b_1,b_2)\leftrightarrow (-b_1,-b_2)$. 
\item At a fixed bolt $\Sigma$, 
the normal space in $M_4$ to any point in $\Sigma$ is a copy of $\R^2$. The orientation 
of this $\R^2$ is fixed from the orientation of $\Sigma$ and the orientation on $M_4$. Near to the bolt we may then similarly 
write $\xi= b_\Sigma\mskip2mu \partial_{\varphi}$, where $\partial_{\varphi}$ rotates 
$\R^2$ with weight $+1$.  Note that equivariant closure of $\Phi$ gives $\ddiff\Phi_0=\xi\hook \Phi_2$, and the 
latter is zero at a fixed point. We deduce that the function $\Phi_0$ is necessarily constant over a connected bolt $\Sigma$. The final term $c_1(N)$ in \eqref{4dBVAB} is the first Chern class of the 
normal bundle $N$ to $\Sigma$ in $M_4$, where the choice of orientation on $\R^2$ makes this into a complex line bundle. In more concrete terms, near to the bolt we may introduce 
plane polar coordinates $(r,\varphi)$ on $\R^2$, and write the metric on $M_4$  to leading order in $r\geq 0$ as
\begin{align}
\ddiff s^2 = \ddiff s^2_\Sigma + \ddiff r^2 +r^2(\ddiff\varphi + \sigma)^2\, .
\end{align}
Here $\sigma$ is a connection one-form for the normal bundle $N$, fibring 
the circle with coordinate $\varphi$ over $\Sigma$.
Then $\int_{\Sigma} c_1(N) = \frac{1}{2\pi}\int_\Sigma \ddiff\sigma\in\mathbb{Z}$ 
describes the twisting of this bundle. 
\end{itemize}

The BVAB formula \eqref{4dBVAB} again 
relates the integral of the top form $\Phi_4$ to the lower degree forms $\Phi_0$, $\Phi_2$
and weights at the fixed points, together with a topological invariant. 
When $M_4$ has a boundary $\partial M_4$, the formula \eqref{4dBVAB} 
has a boundary term added to the right hand side \cite{Couzens:2024vbn}:
\begin{align}\label{bdybvab}
I_{\text{BVAB bdy}}=-\int_{\partial M_4}\frac{1}{\|\xi\|^2}\xi^\flat\wedge \Big(\Phi_2+\Phi_0\frac{\dd\xi^\flat}{\|\xi\|^2}\Big)\, .
\end{align}
Here any metric, for which $\xi$ is a Killing vector, may be used to compute the one-form $\xi^\flat$ and the square norm $\|\xi\|^2$, where we have assumed that 
$\xi\neq 0$ on $\partial M_4$ so that the fixed point set does not intersect the boundary.

\subsection*{Localization of the action}

Our aim in the remainder of this section is to evaluate the 
on-shell action using the equivariantly closed polyform $\Phi$ defined in
\eqref{Phis}, together with the BVAB formula. 

The total action is a sum of the bulk action \eqref{bulkI}
together with boundary terms
\begin{align}\label{4dI}
I =  I_{\text{bulk}}+  I_{\text{bdy}}\, ,
\end{align}
where
\begin{align}\label{boundaryI}
I_{\text{bdy}} =  \frac{1}{16\pi G_4}\bigg[& -\int_{\partial M_4} 
2\mathcal{K} \, \vol_h  \nonumber\\ 
& \qquad +\int_{\partial M_4}(4+R_h)\, \vol_h \bigg]\, .
\end{align}
Here the
boundary $\partial M_4$ has induced metric $h_{ij}$, Ricci scalar $R_h$, trace of extrinsic curvature $\mathcal{K}$, and Riemannian volume form $\vol_h$. 
The first
Gibbons--Hawking--York term in \eqref{boundaryI} is required in order that varying the bulk action \eqref{bulkI}
with respect to a metric fixed on the boundary leads to the 
 Einstein equation in \eqref{EOM}. The second term in \eqref{boundaryI} depends only 
on intrinsic  quantities on the boundary, and is a 
 counterterm that renormalizes the on-shell action 
for asymptotically locally AdS solutions \cite{Emparan:1999pm}. 
More precisely, in \eqref{4dI} one should cut the bulk manifold $M_4$ off 
at some large but finite radius, evaluate the sum of the integrals, and then take the 
cut-off radius to infinity. The result is always finite, 
and independent of the details of the limiting process.

We have already seen that the bulk on-shell action \eqref{bulkIonshell} 
may be expressed as an integral of the top form of the equivariantly closed 
polyform $\Phi$ in \eqref{Phis}. We may then evaluate the full on-shell action \eqref{4dI} using the fixed point formula \eqref{4dBVAB}, 
together with the BVAB boundary term \eqref{bdybvab}. 
It is a remarkable fact that $I_{\text{BVAB bdy}}+
I_{\text{bdy}}=0$ for an \emph{arbitrary} supersymmetric asymptotically  locally AdS
solution. This was shown by a direct calculation in \cite{BenettiGenolini:2019jdz}, 
and generalized to include matter in \cite{BenettiGenolini:2024lbj}. 
This cancellation is surely a direct result of supersymmetry, and one suspects 
there is an argument that does not involve explicit expansions near the boundary of $M_4$. 

The upshot is that the on-shell action receives contributions only from the fixed points
of the supersymmetric Killing vector:
\begin{align}\label{onshellI}
I_{\text{on-shell}} & = \bigg\{\sum_{\text{nuts}\, p}\frac{1}{b_{1,p} b_{2,p}}\Phi_0|_p + 
\sum_{\text{bolts}\, \Sigma}  \int_\Sigma \Big[ \frac{1}{2\pi b_\Sigma}\Phi_2 \nonumber\\
& \qquad \qquad - \frac{1}{b_\Sigma^2}\Phi_0\, c_1(N)\Big]\bigg\}\frac{\pi}{2G_4}\, ,
\end{align}
where $\Phi_0$ and $\Phi_2$ are given in terms of bilinear forms in~\eqref{Phis}. 
The expression  \eqref{onshellI} is very much in the spirit of
 \cite{Gibbons:1979xm}, where holographic renormalization and supersymmetry 
have led to a much sharper result.

\subsection*{Supersymmetry at fixed points}

While \eqref{onshellI} is a striking result, in practice we have 
replaced an integral of $\Phi_4$ by an integral  involving $\Phi_2$ and $\Phi_0$. 
What is remarkable is that these  expressions may be further evaluated in terms of weights and topological invariants of the fixed point set $\mathcal{F}=\{\text{nuts}\}\cup \{\text{bolts}\}$. In order to do this we need to look at the implications of 
supersymmetry at fixed points.

We begin by noting the algebraic (Fierz) identity
\begin{align}
S^2 - P^2 = \|\xi\|^2 \geq 0\, ,
\end{align}
where recall the definitions in \eqref{4dbilinears}. This implies 
\begin{align}
\xi = 0 \quad \Leftrightarrow \quad S = \pm P \quad \Leftrightarrow \quad \gamma_5 \epsilon = \pm \epsilon\, .
\end{align}
The fixed points of $\xi$ are thus precisely points where the Killing spinor 
$\epsilon$ is \emph{chiral}, with $\gamma_5 \epsilon = \pm \epsilon$ referred to as positive/negative chirality. It follows that nuts and bolts have a chirality: $\text{nut}_\pm$, bolt
$\Sigma_\pm$. 

It is next convenient to introduce  self-dual and anti-self-dual projections 
of two-forms: 
\begin{align}
F_\pm \equiv \frac{1}{2}(F\pm *F)\,,
\end{align} with a similar definition for other two-forms.
It follows that $F=F_++F_-$, $* F = F_+ - F_-$.
 The equation for $\ddiff\xi^\flat$ in \eqref{dforms} 
then has anti-self-dual/self-dual parts 
\begin{align}\label{dxiflat}
-\frac{1}{2}(\ddiff \xi^\flat)_\mp = \mp U_\mp \mp (S\mp P)F_\mp\, ,
\end{align}
where here and in the following one should be very careful with signs. 
On the other hand, 
at a fixed point of positive/negative chirality 
$S=\pm P$, so that correlating the chirality and self-duality signs the second term in \eqref{dxiflat} is zero. Again, an algebraic (Fierz) 
identity relates $\|U_\mp\|^2\equiv \tfrac{1}{2!}U_{\mp\mu\nu}U_\mp^{\mu\nu}= \tfrac{1}{2}(S\pm P)^2$. Combining \eqref{dxiflat} with \eqref{nutweight} then gives
\begin{align}\label{Phi0nut}
\mbox{At nut$_\pm$} : \ \  \Phi_0 = -SP & = \mp \frac{1}{4}(S\pm P)^2 \nonumber\\
& = \mp \frac{1}{4}(b_1\mp b_2)^2\, .
\end{align}
Here the second equality simply replaces $S=\pm P$.
The value of $\Phi_0$ at an isolated fixed point is thus entirely determined by the 
weights of $\xi$ at the fixed point, together with the chirality of the spinor.

For a bolt $\Sigma$ we can use the same equations, but differently. Since $\ddiff\xi^\flat$ is an 
exact two-form, integrating the equation for $\ddiff\xi^\flat$ in  \eqref{dforms} over a 
compact bolt $\Sigma$ and using Stokes' theorem gives
\begin{align}\label{starU}
 0 = \int_\Sigma\left( *\mskip3mu U + S *F + PF\right) \, .
\end{align}
On the other hand, $\|U_\pm\|^2= \tfrac{1}{2}(S\mp P)^2$ implies that 
on a bolt $\Sigma_\pm$ one has $U_\pm=0$ and hence $U=\mp * U$. 
Equation  \eqref{starU} then allows us to evaluate the integral of $\Phi_2$ in \eqref{Phis}
(which contains $U$ rather than $*\mskip2mu U$):
\begin{align}\label{Phi2F}
\int_{\Sigma_\pm} \Phi_2 = \int_{\Sigma_\pm} 2S F\, ,
\end{align}
where the terms involving $*\mskip2mu F$ simply drop out. 

The final step involves evaluating the flux integral in~\eqref{Phi2F}. 
The analysis in \eqref{Phi0nut} also goes through for a bolt, where one of the weights is zero (the tangent direction to $\Sigma_\pm$) and 
$S^2=b_\Sigma^2/4$ is constant. In  \cite{BenettiGenolini:2019jdz}  the orientation 
conventions are chosen so that $S=-b_\Sigma/2$ (implying that $b_\Sigma<0)$.
 A global analysis of spinors near to 
the bolt then leads to \cite{BenettiGenolini:2019jdz} 
\begin{align}\label{Fintegral}
\frac{1}{2\pi }\int_{\Sigma_\pm} F = -\frac{1}{2}\int_{\Sigma_\pm}\left[c_1(T)\mp c_1(N)\right]\, .
\end{align}
This is a topological argument: near to the bolt the spinor 
$\epsilon_\pm$ is a section of  $T^{1/2}\otimes N^{\mp 1/2}\otimes \mathcal{L}^{1/2}$. Here $T$ and $N$ are the tangent and normal bundles of $\Sigma_\pm$, and the gauge field $2A$ is a connection on the complex line bundle~$\mathcal{L}$. The latter follows from the spinor being charged under the 
gauge field (the second term in~\eqref{SUSY}). Now $\epsilon_\pm$ 
is a section of a complex line bundle over the bolt $\Sigma_\pm$, but we have also seen that its square norm $S=-b_\Sigma/2$ is constant, and hence nowhere-vanishing. This implies that the complex line bundle is trivial and hence has zero first Chern class, which is precisely \eqref{Fintegral}.
Note that  the right hand side of \eqref{Fintegral} is quantized, being in general a half-integer. In particular 
 $\int_{\Sigma_\pm}c_1(T)=2(1-g)$ is the Chern number of the tangent bundle of the Riemann surface $\Sigma_\pm$, written in term of the genus $g\in\mathbb{Z}_{\geq 0}$ of the surface.

\subsection*{Final formula}

Substituting \eqref{Phi0nut}, \eqref{Fintegral} into \eqref{onshellI}  gives
\begin{align}\label{I4dfinal}
I_{\text{on-shell}}  & =   \bigg[ \sum_{\text{nuts}_\pm}\mp \frac{(b_1\mp b_2)^2}{4b_1b_2} \nonumber \\ 
&  + 
\sum_{\text{bolts}\, \Sigma_\pm } \int_{\Sigma_\pm} \frac{1}{2}c_1(T)\mp\frac{1}{4}c_1(N)\bigg]\frac{\pi}{2G_4}\, .
\end{align}
This is the main result of \cite{BenettiGenolini:2019jdz}, which was re-derived using 
equivariant localization in \cite{BenettiGenolini:2023kxp}. 
Notice that as well as being invariant under $(b_1,b_2)\leftrightarrow (b_2,b_1)$, 
the formula is also invariant under 
$(b_1,b_2)\mapsto \lambda (b_1,b_2)$, for any non-zero constant $\lambda$. 
This had to be the case as there is no natural normalization of $\xi$.

The formula \eqref{I4dfinal} manifestly shows that the on-shell action is an ``equivariant topological invariant'', depending only on weights of the supersymmetric Killing 
vector $\xi$ at fixed points, the chirality of the fixed points, and topological invariants. 
This is not special to $D=4$, $\mathcal{N}=2$ 
gauged supergravity:  as we shall review in section~\ref{sec:matter}, a similar 
formula holds also when the theory is coupled to vector multiplet matter. 
Moreover, similar formulae hold with higher derivative corrections 
added \cite{Bobev:2020egg, Bobev:2021oku, Genolini:2021urf}, and for supergravity theories in other dimensions. 

\subsection*{Examples}

We now illustrate the general formula \eqref{I4dfinal} with a number of simple examples. Although we do not need the metric and gauge field 
to evaluate \eqref{I4dfinal}, we shall present these in some of the examples in order to help orient the reader. 

\vskip 0.25cm

\noindent {\bf Euclidean AdS$_4$}: The simplest solution is the vacuum solution, namely 
Euclidean AdS$_4$ with zero gauge field, $A=0$. 
The metric on global AdS$_4$ may be written
\begin{align}\label{AdSmetric}
\ddiff s^2 = \frac{\ddiff r^2}{r^2+1}+ r^2\left(\ddiff\theta^2+\cos^2\theta\mskip2mu \ddiff\varphi_1^2+\sin^2\theta\mskip2mu \ddiff\varphi_2^2\right)\, .
\end{align}
Here $r\geq 0$ is a radial coordinate, with constant $r>0$ slices being three-spheres: 
$\theta\in[0,\frac{\pi}{2}]$, and $\varphi_i$ have period $2\pi$, $i=1,2$. 
The space has topology $\R^4=\R^2\oplus\R^2$, with the Killing vectors $\partial_{\varphi_i}$ rotating each $\R^2_i$ factor. This is a maximally supersymmetric solution, and there is a four-dimensional family of spinors 
solving \eqref{SUSY}. One can check \cite{BenettiGenolini:2019jdz} that one choice leads to the supersymmetric Killing vector
\begin{align}\label{xiAdS}
\xi = \partial_{\varphi_1}+\partial_{\varphi_2}\, .
\end{align}
Here Euclidean AdS$_4$ has isometry group $SO(4,1)$, with the $SO(4)$ subgroup 
acting $\R^4$ in the obvious way. The vector field \eqref{xiAdS} then generates 
the diagonal $U(1)=SO(2)_{\text{diag}}\subset SO(4)$. 
There is a single fixed point of $\xi$ at the origin $\{r=0\}$, which is a nut of negative chirality. 
From \eqref{xiAdS} we read off $b_1=b_2$ at this nut$_-$, and \eqref{I4dfinal} 
gives
\begin{align}\label{IAdS4}
I_{\text{AdS}_4} = \frac{(1+1)^2}{4 \cdot 1 \cdot 1}\frac{\pi}{2G_4} = \frac{\pi}{2G_4}\, .
\end{align}
Using a different Killing spinor one can also construct a supersymmetric Killing vector with 
$\xi = \partial_{\varphi_1}-\partial_{\varphi_2}$, the anti-diagonal $SO(2)$ in $SO(4)$.  Now $b_1=-b_2$, and the nut has 
positive chirality $\mathrm{nut}_+$. The on-shell action, computed from \eqref{I4dfinal}, of course gives the same final answer in~\eqref{IAdS4}.

\vskip 0.25cm

\noindent {\bf Squashed Euclidean AdS$_4$}: Keeping the topology $\R^4=\R^2\oplus\R^2$ the same as the last example, one can imagine 
more general supersymmetric solutions where
\begin{align}
\xi = b_1 \partial_{\varphi_1}+ b_2\partial_{\varphi_2}\, ,
\end{align}
with $b_1$, $b_2$ now arbitrary parameters. The ratio 
$b_1/b_2$ (or its square root) is often referred to as the ``squashing parameter'' 
\cite{Hama:2011ea}, since for the various solutions cited below 
the metric on the $S^3$ radial slices is squashed/deformed, compared 
to the round three-spheres in the AdS$_4$ metric~\eqref{AdSmetric}. 
 Taking the origin to be a $\mathrm{nut}_-$, we may simply write down the on-shell action:
\begin{align}\label{Isquashed}
I_{\text{on-shell}}= \frac{(b_1+b_2)^2}{4b_1b_2}\frac{\pi}{2G_4}\, .
\end{align}
This precisely matches the on-shell action of the explicit solutions found in
\cite{Martelli:2011fu}, where an anti-self-dual instanton gauge field $F=\ddiff A$ may be turned on while 
preserving the AdS$_4$ metric. These solutions were 
then further generalized in \cite{Martelli:2011fw, Martelli:2012sz, Martelli:2013aqa, Farquet:2014kma}. There also exist solutions with a $\mathrm{nut}_+$ at the origin, 
where the prefactor is $-(b_1-b_2)^2/4b_1b_2$. However, for the anti-self-dual 
solutions in \cite{Farquet:2014kma}, where the Weyl tensor is taken to be anti-self-dual, 
it is shown  that non-singular solutions 
can exist only if certain inequalities on the parameters $b_1,b_2$ hold. 
This is an important point: localization allows one 
to compute various quantities assuming a solution exists, but existence 
is a separate question. In particular, it is natural to ask whether existence imposes simple 
necessary conditions on the supersymmetric Killing vector $\xi$. 

\vskip 0.25cm

\noindent {\bf Hyperbolic black hole}: Our final example in this section 
is the static supersymmetric black hole studied in
\cite{Romans:1991nq, Brill:1997mf, Caldarelli:1998hg}. The metric and gauge field 
in Euclidean signature are
\begin{align}\label{HBH}
\ddiff s^2 & = V(r)\ddiff \tau^2 +\frac{\ddiff r^2}{V(r)} + r^2\ddiff s^2_{\Sigma_g}\, , \nonumber\\ V(r) & =  -1 + r^2 + \frac{\frac{1}{4}-Q^2}{r^2}\, ,\nonumber\\
  A  & = \frac{Q}{r}\ddiff \tau + \frac{1}{2}A_{\Sigma_g}\, .
\end{align}
Here $\ddiff s^2_{\Sigma_g}$ is a constant curvature metric on a Riemann surface of genus $g>1$, 
and $\ddiff A_{\Sigma_g}=\vol_{\Sigma_g}$ is the volume form on $\Sigma_g$. 

Taking the electric charge $Q>0$, the largest root $V(r_0)=0$ is $r_0=\sqrt{\frac{1}{2}+Q}$. For $r\geq r_0$ the solution  then has topology $\R^2\times \Sigma_g$, 
with $\tau$ an angular coordinate on $\R^2$ in plane polars. The supersymmetric Killing vector is (necessarily) $\xi=\partial_\tau$, so that the fixed point set $\mathcal{F}=\{r=r_0\}\cong 
\Sigma_g$ is a bolt, precisely the horizon of the black hole. The magnetic flux through this horizon is computed to be
\begin{align}\label{FBH}
\frac{1}{2\pi}\int_{\Sigma_g}  F = g-1\, ,
\end{align}
in agreement with the general formula \eqref{Fintegral}, where here the normal bundle $N$  is trivial.

Further analysis of the solution may be found in the original references and 
\cite{BenettiGenolini:2019jdz}, but the advantage of supergravity localization
is that one only needs to know the topology of the solution.
The formula \eqref{I4dfinal} allows us to simply write down
\begin{align}\label{IBH}
I_{\text{BH}} = \frac{\pi}{2G_4}(1-g)\, .
\end{align}
Note that the Lorentzian continuation of the solution, taking $t=\ii\tau$, 
has a naked singularity for $Q\neq 0$, and is extremal in the limit $Q=0$. 
One can regard $Q\neq 0$ in Euclidean signature as a regulator, moving
the horizon to finite distance. 
, the on-shell action \eqref{IBH} is  then independent of $Q$, and  agrees with the expected entropy formula $S_{\text{BH}}=-I_{\text{BH}}$ for a static black hole with zero electric charge. 
The result also agrees with
 \cite{Azzurli:2017kxo}, who compute $I_{\text{BH}}$ by instead turning on a 
finite temperature, which also moves the horizon to finite distance but 
breaks supersymmetry. Taking the extremal limit using this method 
agrees with \eqref{IBH}.

\section{Supergravity coupled to matter}\label{sec:matter}

In this section we couple $D=4$, $\mathcal{N}=2$ minimal gauged supergravity to 
vector multiplet matter, following 
\cite{BenettiGenolini:2024xeo, BenettiGenolini:2024lbj}. 
We expect it to be straightforward to include hypermultiplets, 
as was done in a near-horizon setting in~\cite{BenettiGenolini:2024kyy}. 
We will be more brief in this section, highlighting the main formulae 
and new conceptual points.

\subsection*{The theory}

We couple the minimal theory of the last section to an additional $n$ Abelian gauge fields,
so that in total we have $A^I$, $I=0,1,\dots n$, with field strengths $F^I=\dd A^I$. Setting $n=0$ returns us to the minimal theory, with $A^0=A$.
 These vector multiplets also contain 
complex scalar fields $L^I$, where in Euclidean signature the complex conjugate 
$\bar{L}^I\mapsto\widetilde{L}^I$ is an independent field. 
There is some redundancy in this description:  simultaneously rescaling the 
scalar fields by an arbitrary positive function leaves the theory invariant. 
The theory is specified completely by the choice of a prepotential 
$\mathcal{F}=\mathcal{F}(L^I)$, which is an arbitrary homogeneous degree two function of the scalars $L^I$, and constant real Fayet--Iliopoulos (FI) gauging parameters $\xinew_I\in\R$. 
The superpotential and its tilded cousin~are
\begin{align}
W \equiv \zeta_IL^I\, , \quad \widetilde{W} \equiv \zeta_I \widetilde{L}^I\, .
\end{align}

The full Euclidean action may be  found in 
\cite{Bobev:2020pjk, BenettiGenolini:2024xeo, BenettiGenolini:2024lbj}. This includes 
a potential $\mathcal{V}$ for the scalar fields, as well as kinetic terms for the 
scalars and Maxwell fields. 
There is a generalization of the Killing spinor equation in \eqref{SUSY}, 
and an additional algebraic equation for $\epsilon$. 
In particular we may define the bilinear forms exactly as in \eqref{4dbilinears}, 
with the algebraic (Fierz) identities for these hence being the same as in the last section. 

\subsection*{Equivariantly closed forms}

One proceeds in the same way as for minimal supergravity, using the 
relevant Killing spinor and algebraic equation for $\epsilon$. In particular one 
can take covariant derivatives of the bilinears defined in \eqref{4dbilinears}, 
obtaining generalizations of the differential equations \eqref{dforms}. 
One finds that $\xi$ is again a Killing vector that generates a symmetry of the full solution.

Using these differential equations 
we may then similarly construct equivariantly closed forms. For the 
field strengths $F^I=\ddiff A^I$ one finds that
\begin{align}\label{eqvartFI}
\Phi^{I} & = \Phi^I_2+\Phi^I_0 \nonumber\\
& \equiv F^I+\sqrt{2}\left[L^I(S-P)-\widetilde{L}^I(S+P)\right]\, ,
\end{align}
are equivariantly closed, $\ddiff_\xi \Phi^I=0$. Here minimal gauged supergravity is recovered by setting $n=0$ and $L^0=\widetilde{L}^0=\frac{1}{2\sqrt{2}}$, 
so that \eqref{eqvartFI} reduces to $\Phi^F$ in \eqref{Phis}. 
For the on-shell action, $\Phi$ in \eqref{Phis} generalizes to
\begin{equation}
    \Phi=\Phi_4+\Phi_2+\Phi_0\, ,
\end{equation}
where 
\begin{align}\label{phi420}
\Phi_4&\equiv  -\frac{1}{2}\mathcal{V}\mskip2mu \mathrm{vol}_{4}-\frac{1}{4}\mathcal{I}_{IJ}F^I\wedge*F^J+\frac{\ii}{4}\mathcal{R}_{IJ}F^I\wedge F^J\,,\nn\\
\Phi_2
&\equiv    \frac{1}{\sqrt{2}}(WU_++\widetilde{W}U_-) 
 \nonumber\\ 
& \quad  -\frac{1}{\sqrt{2}}\mathcal{I}_{IJ}\big[L^I (S-P)F^{J}_{+} +\widetilde{L}^I (S+P)F^{J}_{-}\big] \nonumber\\
& \quad +\frac{\ii}{\sqrt{2}}\mathcal{R}_{IJ}F^J\big[L^I (S-P)-\widetilde{L}^I (S+P)\big]\,,\nn\\
\Phi_0&\equiv   \ii\big[(S-P)^2\mathcal{F}(L)-(S+P)^2\mathcal{F}(\widetilde{L}) 
\nonumber\\
& \quad +\tfrac{1}{2}(S^2-P^2)\big(\partial_I\mathcal{F}(\widetilde{L})L^I-
 \partial_I\mathcal{F}(L)\widetilde{L}^I\big)\big]\,.
\end{align}
Here $\partial_I\mathcal{F}$ denotes a partial derivative of the prepotential $\mathcal{F}$ 
with respect to its arguments. Notice the self-dual and anti-self-dual projections $U_\pm$, 
$F^I_\pm$, which appeared also in minimal supergravity. The only objects 
we have not defined are $\mathcal{I}_{IJ}$, $\mathcal{R}_{IJ}$, which enter the 
action for the Maxwell fields (the second and third terms in $\Phi_4$).  These are likewise determined entirely by the prepotential, where the formulae in  Euclidean signature may be found in 
 \cite{BenettiGenolini:2024xeo, BenettiGenolini:2024lbj}.
Taking the trace of the Einstein equation and substituting back into the bulk action, one again finds
\begin{align}\label{POSactdef}
I_{\text{bulk, on-shell}} =\left[\frac{1}{(2\pi)^2}\int_{M_4} \Phi_4\right]\frac{\pi}{2G_4} \,.
\end{align}
One can thus evaluate this using  BVAB \eqref{4dBVAB}, being careful with boundary terms.

\subsection*{Localization}

The total action is again a sum of the bulk action \eqref{POSactdef}
together with boundary terms that generalize \eqref{boundaryI}. 
Unlike for minimal gauged supergravity, one can now construct \emph{finite} boundary counterterms, and it is important to choose the coefficients of these so that the 
regularization scheme is  supersymmetric. 
Moreover, as discussed in \cite{Bobev:2020pjk} and appendix~E of \cite{BenettiGenolini:2024lbj}, due to the presence of the scalars,  in order to obtain a supersymmetric observable
one needs to take a Legendre transform of the on-shell action, as a function of certain boundary data for the scalars. We refer to this Legendre transform of the on-shell action as the ``gravitational free energy.''
After a considerable amount of work expanding the equations near the conformal boundary 
of $M_4$, one can show that the boundary 
terms for this quantity all exactly cancel, for a general supersymmetric solution \cite{BenettiGenolini:2024lbj}! It is an important 
question to understand conceptually why this happens, as discussed 
in the last section.
The upshot is the remarkable fixed point formula for the gravitational free energy:
\begin{align}\label{Fgrav}
\Fgrav &  =  \Bigg\{ \sum_{\mathrm{nuts}_\pm}\mp \frac{(b_1\mp b_2)^2}{b_1b_2}\ii \cF(u^J_\pm)  \nonumber\\
& \ \  + \sum_{\mathrm{bolts}\, \Sigma_\pm} \Big[-\pmu \mskip2mu \partial_I \ii\cF(u^J_\pm) \frac{1}{4\pi}\int_{\Sigma_\pm}F^I \nonumber\\
& \qquad \qquad \qquad
\pm \ii \cF(u^J_\pm) \int_{\Sigma_\pm} c_1(N) \Big]\Bigg\}\frac{\pi}{G_4}\, .
\end{align}
Notice immediately that this resembles the minimal result \eqref{I4dfinal}, 
suitably ``dressed'' by the potential $\ii\cF$. In \eqref{Fgrav} the latter is evaluated 
on the following combinations of scalars at fixed points
\begin{align}\label{uI}
\left.u^I_+ \equiv \frac{\widetilde{L}^I}{\sum_{J=0}^n\xinew_J \widetilde{L}^J}\right|_+\, , \quad 
\left.u^I_- \equiv \frac{{L}^I}{\sum_{J=0}^n\xinew_J {L}^J}\right|_-\, .
\end{align}
Thus only the $L^I$ contribute at negative chirality fixed points, while 
only $\widetilde{L}^I$ contribute at positive chirality fixed points. 
By construction note that $\sum_{I=0}^n \zeta_Iu^I_\pm = 1$. 
The constant $\kappa=\pm 1$ is a convention-dependent sign, which is related to the choice of convention 
mentioned above equation \eqref{Fintegral}. The analogue of the latter is 
\begin{align}\label{FIsum}
\frac{1}{4\pi}\int_{\Sigma_\pm} \frac{1}{2}\zeta_IF^I =- \frac{\kappa}{2} \int_{\Sigma_\pm}c_1(T)\mp c_1(N)\, ,
\end{align}
with its derivation identical to that before.
The result for minimal supergravity is recovered on setting $\ii \mathcal{F}(u)=\frac{1}{8}$, $\ii\partial_u\mathcal{F}(u)=1$, where $u=u^0=\tfrac{1}{4}$, $\zeta_0=4$, 
and one takes $\kappa=+1$. 
There is an analogous formula to \eqref{Phi0nut} in the case with matter, 
and an alternative approach to proving this is by considering the 
Lie derivative $\mathcal{L}_\xi \epsilon$ of the spinor along $\xi$, and restricting 
to fixed points. Since we will not need this for the examples below we refer the reader to \cite{BenettiGenolini:2024xeo, BenettiGenolini:2024lbj} for the details.

We note that the localization formula \eqref{Fgrav} gives a general proof of the ``gravitational block'' formulae for certain classes of black holes in \cite{Hosseini:2019iad}, with the ``gluing rules'' 
arising from localizing the integrals of $F^I$ using the equivariant forms in \eqref{eqvartFI}. To see this simply requires applying the formula to rotating 
black hole solutions, as discussed in \cite{BenettiGenolini:2024lbj}.

\subsection*{Examples}

We focus on the STU model, where  $n=3$ and for which the prepotential is
\begin{align}
\ii \mathcal{F}(L^I) = 2\sqrt{L^0L^1L^2L^3}\, .
\end{align}
The FI parameters $\zeta_I=1$ for all $I=0,1,2,3$. Solutions to this theory 
uplift on the seven-sphere $S^7$ to solutions of $D=11$ supergravity \cite{Cvetic:1999xp}, and are holographically dual to the ABJM theory on 
the conformal boundary $\partial M_4$.

\vskip 0.25cm

\noindent {\bf Black saddle solutions}: Our first example is a generalization 
of the hyperbolic black hole \eqref{HBH} to include matter. 
The terminology ``black saddle''  means these should be viewed 
as (putative) saddle points to the Euclidean gravitational path integral. In particular, unless 
$\widetilde{L}^I=\bar{L}^I$ these solutions do not have good Lorentzian continuations, just 
as for the hyperbolic black hole with $Q\neq 0$. 
Various analytic and numerical solutions were found in~\cite{Bobev:2020pjk}.
The beauty of the general formula \eqref{Fgrav} is that this allows us 
to recover the results for the gravitational free energy, without any details 
of the solutions, but assuming they exist.

As for the hyperbolic black hole, we consider solutions with topology
$M_4=\R^2\times \Sigma_g$. The supersymmetric Killing vector $\xi=\partial_\tau$ 
again rotates the $\R^2$ factor fixing the origin, which is the horizon of the black saddle. 
We may define quantized magnetic charges
\begin{align}
p^I \equiv \frac{1}{4\pi}\int_{\Sigma_g} F^I\, .
\end{align}
Due to \eqref{FIsum}, and setting $\kappa=+1$, these are constrained to satisfy
\begin{align}
\sum_{I=0}^3 p^I = 2g-2\, .
\end{align}
There is a single bolt, and the normal bundle $N$ is trivially fibred, so only the second term in 
\eqref{Fgrav} contributes. We write down
\begin{align}\label{Fgravsaddles}
\Fgrav= -2\sqrt{u^0u^1u^2u^3} \sum_{I=0}^3 \frac{p^I}{u^I}\frac{\pi}{2G_4}\, .
\end{align}
Here the formula is formally the same for both chiralities of the bolt, where $u^I=u^I_\pm$ 
respectively, and $\sum_{I=0}^3 u^I=1$ follows from the definitions \eqref{uI}. 

In \eqref{Fgrav} and \eqref{Fgravsaddles} the scalars $u^I_\pm$ are evaluated
 on the fixed point set. However, in holography one would like to express 
quantities as functions of data on the conformal boundary 
$\partial M_4$. To see how this works for the gravitational free energy for the black saddles, consider now integrating $F^I$ over the $\R^2$ direction,
at some choice of point on the horizon $\Sigma_g$. One can always choose a gauge 
so that on this $\R^2$ we have $F^I=\ddiff A^I$ for a global one-form $A^I$. 
Then define
\begin{align}
\Delta^I \equiv \frac{1}{4\pi}\int_{\R^2}F^I  = \frac{1}{4\pi}\int_{S^1} A^I\, .
\end{align}
Here $S^1$ is the boundary of $\R^2$ on the UV conformal boundary 
$\partial M_4 = S^1\times \Sigma_g$. Likewise, we can define the boundary value
 \begin{align}
\sigma^I \equiv -\frac{\ii}{4\pi}\Phi^I_0|_{S^1}\, ,
\end{align}
where $\Phi^I_0$ is given in \eqref{eqvartFI}.  Then localization on
the $\R^2$ gives the ``IR-UV'' relation
\cite{BenettiGenolini:2024xeo, BenettiGenolini:2024lbj} 
\begin{align}\label{UVIR}
u^I = \Delta^I+\ii\beta\mskip1mu\sigma^I\, ,
\end{align}
where $\beta$ is the period of the Euclidean time circle $\tau$, with $\xi=\partial_\tau$. 
The quantities on the right hand side of \eqref{UVIR} are defined 
on the conformal boundary $\partial M_4 = S^1\times \Sigma_g$, and 
are thus part of the data that specifies the background that the dual 
supersymmetric conformal theory is defined on. The left hand side 
of \eqref{UVIR} are the scalars at the horizon, which enter \eqref{Fgravsaddles}.
Thus importantly \eqref{UVIR} allows us to interpret the gravitational free energy \emph{as a function of boundary data}. Although we have described this for a particular example, 
this generalizes for solutions of more general topology  -- see \cite{BenettiGenolini:2024hyd} for 
quite explicit formulae. Moreover, with this interpretation, \eqref{Fgravsaddles}
precisely matches the large $N$ limit of
the topologically twisted index of the ABJM theory on 
$S^1\times \Sigma_g$ \cite{Bobev:2020pjk} (for $g\neq 1$). 

\vskip 0.25cm

\noindent {\bf Black bolt solutions}: It is straightforward to generalize 
the above analysis to the case that the $\R^2$ factor is fibred 
over the horizon $\Sigma_g$. This means that the Chern number
\begin{align}
p\equiv - \int_{\Sigma_g}{c_1(N)}\in\mathbb{Z}\, ,
\end{align}
is non-zero. Solutions with this type of fibred topology are known
as ``Taub-bolt'' solutions in the gravity literature. Such supersymmetric solutions to
the Euclidean STU theory are not known, although there
are explicit solutions in minimal supergravity constructed
in \cite{Martelli:2012sz, Toldo:2017qsh}. From \eqref{Fgrav} we can again write down,
for positive/negative chirality bolts,
\begin{align}\label{Fgravbolts}
\Fgrav= -2\sqrt{u^0u^1u^2u^3}\left(\pm 2p+ \sum_{I=0}^3 \frac{p^I}{u^I}\right)\frac{\pi}{2G_4}\, ,
\end{align}
where the constraint on the magnetic fluxes \eqref{FIsum} is now
\begin{align}\label{pIp}
\sum_{I=0}^3 p^I = \mp p + 2g-2\, .
\end{align}
As first pointed out in \cite{BenettiGenolini:2024xeo}, 
\eqref{Fgravbolts}, \eqref{pIp} precisely match the large $N$ field theory result 
computed in \cite{Toldo:2017qsh}. 

This is an example where 
supergravity localization allows one to compute a gravitational result, 
and match it to field theory, when the gravity solutions are not actually known. 
More importantly, localization allows us to compute explicit results for solutions that are very unlikely to ever be constructed in 
closed form. A good example is the toric class of solutions discussed in 
\cite{BenettiGenolini:2024hyd}.

\section{Discussion}\label{sec:discuss}

We close with some questions, comments, and directions for future work:

\begin{itemize}
\item  From the growing body of work referenced in the introduction, it is 
now clear that equivariant forms are a general feature of supergravity theories.
However, is there a general structure and construction? For example, can 
equivariantly closed forms be built from components of supermultiplets in a systematic way? 

\item Is there a ``protected'' sector of supergravity theories that equivariant 
forms are capturing? Can one define a  cohomological theory?
 What other observables or invariants 
can be computed? Is there a relation between the structure described here and the twisted holography programme  \cite{Costello:2018zrm}? 

\item The formulae we have derived assume that solutions exist. 
In turn, existence is a hard PDE problem. One might envisage proving 
some simple necessary conditions for existence, based on physical
and/or mathematical arguments.  The only general results of which we are aware are discussed in \cite{Farquet:2014kma, BenettiGenolini:2024kyy}. Reference  \cite{Farquet:2014kma}
 studied the class of anti-self-dual solutions 
to minimal gauged supergravity, where the Weyl tensor is assumed to be 
anti-self-dual. For solutions with topology $M_4=\R^4$, with a single nut at the origin (see equation \eqref{Isquashed}), certain constraints on the allowed values of the weights $b_1,b_2$ at the nut
were derived for \emph{non-singular} solutions. This used the differential equations  in an essential way. Instead in \cite{BenettiGenolini:2024kyy} (and for the uplifted solutions in \cite{Couzens:2018wnk}) positivity 
of the central charge and scaling dimensions rule out certain solutions.
These arguments presumably generalize, and it would be interesting to investigate this further.

\item We have focused on two-derivative supergravity, but as already mentioned 
the result \eqref{I4dfinal} is  known to extend to four derivatives  \cite{Bobev:2020egg, Bobev:2021oku, Genolini:2021urf}. It would be interesting if this could be proven using localization, along
with the more general conjectures concerning all-deriviative results
 \cite{Hristov:2021qsw, Hristov:2022plc}. 
Most ambitiously, one might hope to localize the Euclidean supergravity 
path integral itself,  developing the ideas of \cite{Dabholkar:2011ec, Dabholkar:2014wpa}. 

\item As pointed out in \cite{BenettiGenolini:2024lbj}, the results we obtain 
for on-shell actions are in agreement with large $N$ field theory results 
even when there is no consistent Kaluza--Klein truncation. Specifically, 
this is the case for  general classes of three-dimensional $\mathcal{N}=2$ superconformal field theories \cite{Toldo:2017qsh} that
have $D=11$ holographic duals involving a Sasaki--Einstein seven-manifold internal space. This could be explained by directly localizing in $D=11$. 
We leave this, and other speculations, for future work.

\end{itemize}

\section*{Acknowledgments}
This article is based on a talk given at the \emph{Roberto Salmeron School
in Mathematical Physics}, held at the University of Bras\'ilia in September 2025, 
to appear in the proceedings book \emph{Modern topics in mathematical physics}. I would like to thank the organizers of the School, in particular Carolina Matt\'e Gregory, for their generous hospitality, and Chris Couzens, Jerome Gauntlett, Alice L{\"u}scher and Carolina Matt\'e Gregory for comments on the draft. I would also like to thank 
my other localizing collaborators: 
Pietro Benetti Genolini, Yusheng Jiao, Davide Muniz,  Jaeha Park and Tabea Sieper. 
This work is supported in part by STFC grant ST/X000761/1. 

\renewcommand{\bibname}{References} \begingroup
\let\cleardoublepage\relax


\begin{thebibliography}{99}

\bibitem{Benini:2015eyy}
F.~Benini, K.~Hristov and A.~Zaffaroni,
``Black hole microstates in AdS$_{4}$ from supersymmetric localization,''
JHEP \textbf{05} (2016), 054
doi:10.1007/JHEP05(2016)054.

\bibitem{Benini:2016rke}
F.~Benini, K.~Hristov and A.~Zaffaroni,
``Exact microstate counting for dyonic black holes in AdS4,''
Phys. Lett. B \textbf{771} (2017), 462-466
doi:10.1016/j.physletb.2017.05.076.

\bibitem{Pestun:2016zxk}
V.~Pestun, M.~Zabzine, F.~Benini, T.~Dimofte, T.~T.~Dumitrescu, K.~Hosomichi, S.~Kim, K.~Lee, B.~Le Floch and M.~Marino, \textit{et al.}
``Localization techniques in quantum field theories,''
J. Phys. A \textbf{50} (2017) no.44, 440301
doi:10.1088/1751-8121/aa63c1.

\bibitem{Gibbons:1976ue}
G.~W.~Gibbons and S.~W.~Hawking,
``Action Integrals and Partition Functions in Quantum Gravity,''
Phys. Rev. D \textbf{15} (1977), 2752-2756
doi:10.1103/PhysRevD.15.2752

\bibitem{BenettiGenolini:2023kxp}
P.~Benetti Genolini, J.~P.~Gauntlett and J.~Sparks,
``Equivariant Localization in Supergravity,''
Phys. Rev. Lett. \textbf{131}, no. 12, 121602 (2023)
doi:10.1103/PhysRevLett.131.121602.

\bibitem{BenettiGenolini:2024xeo}
P.~Benetti Genolini, J.~P.~Gauntlett, Y.~Jiao, A.~L{\"u}scher and J.~Sparks,
``Localization of the Free Energy in Supergravity,''
Phys. Rev. Lett. \textbf{133} (2024) no.14, 141601
doi:10.1103/PhysRevLett.133.141601.

\bibitem{BenettiGenolini:2024hyd}
P.~Benetti Genolini, J.~P.~Gauntlett, Y.~Jiao, A.~L{\"u}scher and J.~Sparks,
``Toric gravitational instantons in gauged supergravity,''
Phys. Rev. D \textbf{111} (2025) no.4, 046024
doi:10.1103/PhysRevD.111.046024.

\bibitem{BenettiGenolini:2024lbj}
P.~Benetti Genolini, J.~P.~Gauntlett, Y.~Jiao, A.~L{\"u}scher and J.~Sparks,
``Equivariant localization for D = 4 gauged supergravity,''
JHEP \textbf{08} (2025), 211
doi:10.1007/JHEP08(2025)211.

\bibitem{Hristov:2024cgj}
K.~Hristov,
``Equivariant localization and gluing rules in 4d $\mathcal{N}=2$ higher derivative supergravity,''
[arXiv:2406.18648 [hep-th]].

\bibitem{Cassani:2024kjn}
D.~Cassani, A.~Ruip{\'e}rez and E.~Turetta,
``Localization of the 5D supergravity action and Euclidean saddles for the black hole index,''
JHEP \textbf{12} (2024), 086
doi:10.1007/JHEP12(2024)086.

\bibitem{Colombo:2025ihp}
E.~Colombo, V.~Dimitrov, D.~Martelli and A.~Zaffaroni,
``Equivariant localization in supergravity in odd dimensions,''
[arXiv:2502.15624 [hep-th]].

\bibitem{BenettiGenolini:2025icr}
P.~Benetti Genolini, J.~P.~Gauntlett, Y.~Jiao, J.~Park and J.~Sparks,
``Equivariant localization for $D=5$ gauged supergravity,''
[arXiv:2508.08207 [hep-th]].

\bibitem{Colombo:2025yqy}
E.~Colombo, V.~Dimitrov, D.~Martelli and A.~Zaffaroni,
``Patch-wise localization with Chern-Simons forms in five dimensional supergravity,''
[arXiv:2511.13824 [hep-th]].

\bibitem{Park:2025fon}
J.~Park,
``Localizing AlAdS$_5$ black holes and the SUSY index on $S^1 \times M_3$,''
[arXiv:2511.15666 [hep-th]].

\bibitem{Couzens:2025ghx}
C.~Couzens, C.~M.~Gregory, D.~Muniz, T.~Sieper and J.~Sparks,
``Localizing Romans supergravity,''
JHEP \textbf{09} (2025), 103
doi:10.1007/JHEP09(2025)103.

\bibitem{BenettiGenolini:2024kyy}
P.~Benetti Genolini, J.~P.~Gauntlett, Y.~Jiao, A.~L{\"u}scher and J.~Sparks,
``Localization and attraction,''
JHEP \textbf{05} (2024), 152
doi:10.1007/JHEP05(2024)152.

\bibitem{BenettiGenolini:2023ndb}
P.~Benetti Genolini, J.~P.~Gauntlett and J.~Sparks,
``Equivariant localization for AdS/CFT,''
JHEP \textbf{02} (2024), 015
doi:10.1007/JHEP02(2024)015.

\bibitem{BenettiGenolini:2023yfe}
P.~Benetti Genolini, J.~P.~Gauntlett and J.~Sparks,
``Localizing wrapped M5-branes and gravitational blocks,''
Phys. Rev. D \textbf{108} (2023) no.10, L101903
doi:10.1103/PhysRevD.108.L101903.

\bibitem{Suh:2024asy}
M.~Suh,
``Equivariant localization for wrapped M5-branes and D4-branes,''
[arXiv:2404.01386 [hep-th]].

\bibitem{Couzens:2024vbn}
C.~Couzens and A.~L{\"u}scher,
``A geometric dual of F-maximization in massive type IIA,''
JHEP \textbf{08} (2024), 218
doi:10.1007/JHEP08(2024)218.

\bibitem{Couzens:2025nxw}
C.~Couzens, A.~L{\"u}scher and J.~Sparks,
``Localizing punctures in M-theory,''
[arXiv:2511.03397 [hep-th]].

\bibitem{Couzens:2026qne}
C.~Couzens, A.~L{\"u}scher and J.~Sparks,
``IIB an equivariantly localized puncture,''
[arXiv:2601.07598 [hep-th]].

\bibitem{Cassia:2025aus}
L.~Cassia and K.~Hristov,
``Constant maps in equivariant topological strings and geometric modeling of fluxes,''
J. Phys. A \textbf{58} (2025) no.49, 495201
doi:10.1088/1751-8121/ae22ac.

\bibitem{Cassia:2025jkr}
L.~Cassia and K.~Hristov,
``M2-brane partition functions and HD supergravity from equivariant topological strings,''
[arXiv:2508.21619 [hep-th]].

\bibitem{Martelli:2023oqk}
D.~Martelli and A.~Zaffaroni,
``Equivariant localization and holography,''
Lett. Math. Phys. \textbf{114} (2024) no.1, 15
doi:10.1007/s11005-023-01752-1.

\bibitem{Colombo:2023fhu}
E.~Colombo, F.~Faedo, D.~Martelli and A.~Zaffaroni,
``Equivariant volume extremization and holography,''
JHEP \textbf{01} (2024), 095
doi:10.1007/JHEP01(2024)095.

\bibitem{BV:1982}
N.~Berline and M.~Vergne, ``Classes caract\'{e}ristiques \'{e}quivariantes. Formules de local- isation en cohomologie \'{e}quivariante,'' C.R. Acad. Sc. Paris \textbf{295}
 (1982), 539-541

\bibitem{Atiyah:1984px}
M.~F.~Atiyah and R.~Bott,
``The moment map and equivariant cohomology,''
Topology \textbf{23} (1984), 1-28
doi:10.1016/0040-9383(84)90021-1

\bibitem{Freedman:1976aw}
D.~Z.~Freedman and A.~K.~Das,
``Gauge Internal Symmetry in Extended Supergravity,''
Nucl. Phys. B \textbf{120} (1977), 221-230
doi:10.1016/0550-3213(77)90041-4.

\bibitem{Genolini:2021urf}
P.~B.~Genolini and P.~Richmond,
``Supersymmetry of higher-derivative supergravity in AdS4 holography,''
Phys. Rev. D \textbf{104} (2021) no.6, L061902
doi:10.1103/PhysRevD.104.L061902.

\bibitem{Freedman:2013oja}
D.~Z.~Freedman and S.~S.~Pufu,
``The holography of $F$-maximization,''
JHEP \textbf{03} (2014), 135
doi:10.1007/JHEP03(2014)135.

\bibitem{Gauntlett:2007ma}
J.~P.~Gauntlett and O.~Varela,
``Consistent Kaluza-Klein reductions for general supersymmetric AdS solutions,''
Phys. Rev. D \textbf{76} (2007), 126007
doi:10.1103/PhysRevD.76.126007.

\bibitem{Aharony:2008ug}
O.~Aharony, O.~Bergman, D.~L.~Jafferis and J.~Maldacena,
``N=6 superconformal Chern-Simons-matter theories, M2-branes and their gravity duals,''
JHEP \textbf{10} (2008), 091
doi:10.1088/1126-6708/2008/10/091.

\bibitem{Caldarelli:2003pb}
M.~M.~Caldarelli and D.~Klemm,
``All supersymmetric solutions of N=2, D = 4 gauged supergravity,''
JHEP \textbf{09} (2003), 019
doi:10.1088/1126-6708/2003/09/019.

\bibitem{BenettiGenolini:2019jdz}
P.~Benetti Genolini, J.~M.~Perez Ipi{\~n}a and J.~Sparks,
``Localization of the action in AdS/CFT,''
JHEP \textbf{10} (2019), 252
doi:10.1007/JHEP10(2019)252.

\bibitem{Gauntlett:2005bn}
J.~P.~Gauntlett,
``Classifying supergravity solutions,''
Fortsch. Phys. \textbf{53} (2005), 468-479
doi:10.1002/prop.200510206.

\bibitem{Witten:1982im}
E.~Witten,
``Supersymmetry and Morse theory,''
J. Diff. Geom. \textbf{17} (1982) no.4, 661-692

\bibitem{Gibbons:1979xm}
G.~W.~Gibbons and S.~W.~Hawking,
``Classification of Gravitational Instanton Symmetries,''
Commun. Math. Phys. \textbf{66} (1979), 291-310
doi:10.1007/BF01197189

\bibitem{Emparan:1999pm}
R.~Emparan, C.~V.~Johnson and R.~C.~Myers,
``Surface terms as counterterms in the AdS / CFT correspondence,''
Phys. Rev. D \textbf{60} (1999), 104001
doi:10.1103/PhysRevD.60.104001.

\bibitem{Bobev:2020egg}
N.~Bobev, A.~M.~Charles, K.~Hristov and V.~Reys,
``The Unreasonable Effectiveness of Higher-Derivative Supergravity in AdS$_4$ Holography,''
Phys. Rev. Lett. \textbf{125} (2020) no.13, 131601
doi:10.1103/PhysRevLett.125.131601.

\bibitem{Bobev:2021oku}
N.~Bobev, A.~M.~Charles, K.~Hristov and V.~Reys,
``Higher-derivative supergravity, AdS$_{4}$ holography, and black holes,''
JHEP \textbf{08} (2021), 173
doi:10.1007/JHEP08(2021)173.

\bibitem{Hama:2011ea}
N.~Hama, K.~Hosomichi and S.~Lee,
``SUSY Gauge Theories on Squashed Three-Spheres,''
JHEP \textbf{05} (2011), 014
doi:10.1007/JHEP05(2011)014.

\bibitem{Martelli:2011fu}
D.~Martelli, A.~Passias and J.~Sparks,
``The gravity dual of supersymmetric gauge theories on a squashed three-sphere,''
Nucl. Phys. B \textbf{864} (2012), 840-868
doi:10.1016/j.nuclphysb.2012.07.019.

\bibitem{Martelli:2011fw}
D.~Martelli and J.~Sparks,
``The gravity dual of supersymmetric gauge theories on a biaxially squashed three-sphere,''
Nucl. Phys. B \textbf{866} (2013), 72-85
doi:10.1016/j.nuclphysb.2012.08.015.

\bibitem{Martelli:2012sz}
D.~Martelli, A.~Passias and J.~Sparks,
``The supersymmetric NUTs and bolts of holography,''
Nucl. Phys. B \textbf{876} (2013), 810-870
doi:10.1016/j.nuclphysb.2013.04.026.

\bibitem{Martelli:2013aqa}
D.~Martelli and A.~Passias,
``The gravity dual of supersymmetric gauge theories on a two-parameter deformed three-sphere,''
Nucl. Phys. B \textbf{877} (2013), 51-72
doi:10.1016/j.nuclphysb.2013.09.012.

\bibitem{Farquet:2014kma}
D.~Farquet, J.~Lorenzen, D.~Martelli and J.~Sparks,
``Gravity duals of supersymmetric gauge theories on three-manifolds,''
JHEP \textbf{08} (2016), 080
doi:10.1007/JHEP08(2016)080.

\bibitem{Romans:1991nq}
L.~J.~Romans,
``Supersymmetric, cold and lukewarm black holes in cosmological Einstein-Maxwell theory,''
Nucl. Phys. B \textbf{383} (1992), 395-415
doi:10.1016/0550-3213(92)90684-4.

\bibitem{Brill:1997mf}
D.~R.~Brill, J.~Louko and P.~Peldan,
``Thermodynamics of (3+1)-dimensional black holes with toroidal or higher genus horizons,''
Phys. Rev. D \textbf{56} (1997), 3600-3610
doi:10.1103/PhysRevD.56.3600.

\bibitem{Caldarelli:1998hg}
M.~M.~Caldarelli and D.~Klemm,
``Supersymmetry of Anti-de Sitter black holes,''
Nucl. Phys. B \textbf{545} (1999), 434-460
doi:10.1016/S0550-3213(98)00846-3.

\bibitem{Azzurli:2017kxo}
F.~Azzurli, N.~Bobev, P.~M.~Crichigno, V.~S.~Min and A.~Zaffaroni,
``A universal counting of black hole microstates in AdS$_{4}$,''
JHEP \textbf{02} (2018), 054
doi:10.1007/JHEP02(2018)054.

\bibitem{Bobev:2020pjk}
N.~Bobev, A.~M.~Charles and V.~S.~Min,
``Euclidean black saddles and AdS$_{4}$ black holes,''
JHEP \textbf{10} (2020), 073
doi:10.1007/JHEP10(2020)073.

\bibitem{Hosseini:2019iad}
S.~M.~Hosseini, K.~Hristov and A.~Zaffaroni,
``Gluing gravitational blocks for AdS black holes,''
JHEP \textbf{12} (2019), 168
doi:10.1007/JHEP12(2019)168.

\bibitem{Cvetic:1999xp}
M.~Cvetic, M.~J.~Duff, P.~Hoxha, J.~T.~Liu, H.~Lu, J.~X.~Lu, R.~Martinez-Acosta, C.~N.~Pope, H.~Sati and T.~A.~Tran,
``Embedding AdS black holes in ten-dimensions and eleven-dimensions,''
Nucl. Phys. B \textbf{558} (1999), 96-126
doi:10.1016/S0550-3213(99)00419-8.

\bibitem{Toldo:2017qsh}
C.~Toldo and B.~Willett,
``Partition functions on 3d circle bundles and their gravity duals,''
JHEP \textbf{05} (2018), 116
doi:10.1007/JHEP05(2018)116.

\bibitem{Costello:2018zrm}
K.~Costello and D.~Gaiotto,
``Twisted holography,''
JHEP \textbf{01} (2025), 087
doi:10.1007/JHEP01(2025)087.

\bibitem{Couzens:2018wnk}
C.~Couzens, J.~P.~Gauntlett, D.~Martelli and J.~Sparks,
``A geometric dual of $c$-extremization,''
JHEP \textbf{01} (2019), 212
doi:10.1007/JHEP01(2019)212.

\bibitem{Hristov:2021qsw}
K.~Hristov,
``4d $ \mathcal{N} $ = 2 supergravity observables from Nekrasov-like partition functions,''
JHEP \textbf{02} (2022), 079
doi:10.1007/JHEP02(2022)079.

\bibitem{Hristov:2022plc}
K.~Hristov,
``Maximally symmetric nuts in 4d $\mathcal{N} = 2$ higher derivative supergravity,''
JHEP \textbf{02} (2023), 110
doi:10.1007/JHEP02(2023)110.

\bibitem{Dabholkar:2011ec}
A.~Dabholkar, J.~Gomes and S.~Murthy,
``Localization {\&} Exact Holography,''
JHEP \textbf{04} (2013), 062
doi:10.1007/JHEP04(2013)062.

\bibitem{Dabholkar:2014wpa}
A.~Dabholkar, N.~Drukker and J.~Gomes,
``Localization in supergravity and quantum $AdS_4/CFT_3$ holography,''
JHEP \textbf{10} (2014), 090
doi:10.1007/JHEP10(2014)090.

\end{thebibliography}
\end{document}